\documentclass[aps,prl,twocolumn,showpacs,amsmath,amssymb]{revtex4}
\usepackage{epsfig}
\usepackage{bm}
\newcommand{\av}[1]{\left\langle{#1}\right\rangle}

\begin{document}
\title{Fluctuation relations without microreversibility in nonlinear transport}
\author{H. F\"orster and  M. B\"uttiker} 
\affiliation{D\'epartement de 
Physique Th\'eorique, Universit\'e de Gen\`eve, CH-1211 Gen\`eve 4,
Switzerland}

\date{\today}

\begin{abstract}
In linear transport, the fluctuation-dissipation theorem relates 
equilibrium current correlations to the linear conductance coefficient. 
For nonlinear transport, there exist fluctuation relations that rely on
Onsager's principle of microscopic 
reversibility away from equilibrium. However, both theory and experiments have
shown  
deviations from microreversibility in the form of magnetic field asymmetric  
current-voltage relations. 
We present novel fluctuation relations for nonlinear transport in the presence
of  magnetic fields that
relate current correlation functions at any order at equilibrium to response
coefficients of current cumulants of lower order. We illustrate our results 
with the example of an electrical Mach-Zehnder interferometer. 
\end{abstract}

\maketitle

{\it Introduction} --
Onsager derived the symmetry of transport coefficients of irreversible
processes using the principle of microscopic reversibility for the
fluctuations of the equilibrium system \cite{Onsager1,casimir}. 
Thus the symmetry of transport 
coefficients in the linear transport regime is directly related to the
fluctuation-dissipation theorem of Einstein, Johnson, Nyquist and Kubo \cite{einstein,kubo,johnson,nyquist}. 
Naturally, the question arises whether there are fluctuation relations
beyond the linear response regime. In statistical mechanics,  
fluctuation relations
were derived \cite{evans, gallavotti} as an extension of Onsager's relation
to systems far from equilibrium. These fluctuation relations
make statements on the distribution function of observables
conjugate to thermodynamic forces for 
a wide variety of non-equilibrium systems \cite{taniguchi, rondoni}. 
In electrical transport the variable of interest is the transferred charge. 
The theory is known as full counting statistics \cite{levitov93,tobiska}, and
fluctuation relations for conductors have been discussed in the absence
 \cite{tobiska, andrieux, espositoPRB,astumian} and the presence of  a
 magnetic field  \cite{saito}.

At equilibrium, in the presence of a magnetic field, Onsager reciprocity still
holds. However, away from equilibrium, the potential landscape inside the
conductor is neither an odd nor an even function of magnetic field. As a
consequence, electrical conductors exhibit manifest deviations from symmetries
based on microreversibility and fluctuation relations derived from this
principle \cite{saito} are not valid.   Surprisingly, and this is a central
point of our work, we obtain novel fluctuation relations even without invoking
the principle of microreversibility. Importantly, the novel fluctuation
relations are general and independent of a specific model for interactions.

\medskip
{\it Full counting statistics and fluctuation theorem} -- 
The full counting statistics of a conductor with $M$ terminals is
the probability distribution 
$P({\bf Q})$ that ${\bf Q}=(Q_1,Q_2,\ldots,Q_M)$ charges are
transmitted into the  reservoirs during the measurement time
$t$. The distribution function $P({\bf Q})$ is expressed by the generating
function $F(i{\bf \Lambda})=\ln\sum_{{\bf Q}}P({\bf Q})e^{i{\bf \Lambda Q}}$,
where ${\bf 
  \Lambda}=(\lambda_1,\lambda_2,\ldots,\lambda_M)$ are  called counting 
fields. 
In the long time limit, all irreducible current cumulants at zero frequency
are obtained by consecutive derivatives of the generating function, in contact
$\alpha$ this is $\av{(\Delta I_\alpha)^k}= (-i e)^k[\partial^k F/\partial
  \lambda_\alpha^k]_{\bf   \Lambda=0}/t$.   
The magnetic field $B$ perpendicular to the conductor and the
affinities   ${\bf
A}=(\frac{eV_1}{k_BT},\frac{eV_2}{k_BT},\ldots,\frac{eV_M}{k_BT})$ are externally controlled. 
Here 
$eV_\alpha$ is the potential at terminal $\alpha$ and $T$ the temperature,
assumed to be equal and non-zero in all terminals.

\begin{figure}[t]   
\centerline{\psfig{figure=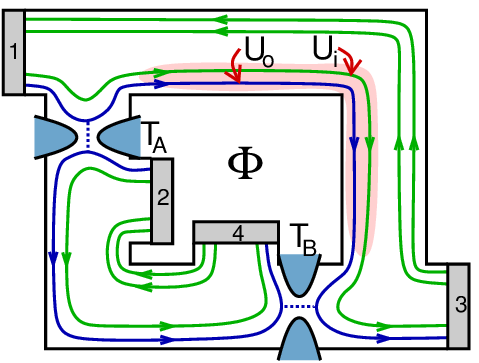,width=8.0cm}}   
\caption{(color online) Mach-Zehnder interferometer at filling factor
  $2$. Only the outer 
  edge  state enters the interferometer. Coulomb-interactions between
  neighboring edges -as indicated by shading- lead to internal potentials
  $U_i$ and $U_o$ in the inner and outer edge.  For
  reversed magnetic field, all  arrows point in the opposite direction.}
\label{mzi}
\end{figure}

The fluctuation relation  for the full counting
statistics gives a simple relation for the probability that ${\bf Q}$ 
or -at reversed magnetic field- ${\bf-Q}$ charges are transmitted. 
Derivations \cite{tobiska,andrieux,espositoPRB,astumian,saito} rely
fundamentally on microscopic reversibility: 
a process from terminal $\alpha$ to $\beta$ has the same probability as  the
reversed process, from terminal $\beta$ to $\alpha$ at inversed magnetic 
field. Refs.~\cite{tobiska,andrieux,espositoPRB,astumian,saito} assume that
this is valid also far from equilibrium and find
\begin{eqnarray}
	P({\bf Q},B)&=&e^{{\bf A Q}}P(-{\bf Q},-B)\label{FTfcs}\\ 
	F_\pm(i{\bf\Lambda, A})&=&\pm F_\pm(-i{\bf
	\Lambda-A,A})\label{sympm} 
\end{eqnarray}
Eq.~(\ref{sympm}) is the Fourier transform of Eq.~(\ref{FTfcs}) and determines
the symmetry of the generating function. For convenience, the
\mbox{(anti-)}symmetrized generating function 
$F_\pm(i{\bf\Lambda})=F(i{\bf\Lambda},B)\pm F(i{\bf\Lambda},-B)$ is used, and
the  notation $F_\pm(i{\bf\Lambda,A})$ emphasizes that the
generating function depends explicitly on the affinities ${\bf A}$.
The derivation \cite{tobiska,andrieux,espositoPRB,saito} of Eq.~(\ref{sympm})
relies 
fundamentally on microscopic reversibility: 
a process from terminal $\alpha$ to $\beta$ has the same probability as  the
reversed process, from terminal $\beta$ to $\alpha$ at inversed magnetic 
field (see appendix A). 

\medskip
{\it The Mach-Zehnder interferometer (MZI)} --
As an instructive example, we present a MZI (see
Fig. 1) and show how interaction (screening) effects lead to deviations from
reversibility.
It is a four terminal conductor with two quantum point contacts (QPC) acting
as  beam splitters as shown in Fig.~\ref{mzi}. The two interferometer arms
enclose 
a magnetic flux $\Phi$, such that interference arises due to the Aharonov-Bohm
effect. In experiments \cite{ji, strunk,roulleau, bieri}, the MZI is realized
using edge states in the quantum 
Hall regime, and it is often operated at filling factor $2$. Only
carriers in the outer edge enter the interferometer and are able to
interfere. Here, the inner edge state  moves in vicinity in both interferometer
arms and carries current from terminal $1$ to $3$ and from $2$ to $4$
 \cite{bieri}. Although a four-terminal conductor, the MZI is
characterized by only a single transmission probability $T_{31}$, due to the
separation of  left- and right movers. $T_{31}$ is the probability for a
particle to be transmitted in the outer edge state from terminal $1$ to
$3$. In the linear transport regime reciprocity means that \cite{buttikeribm88}
$T_{31}(+B)=T_{13}(-B)$. We next demonstrate that already Hartree interactions
lead to a violation of  Eqs.~(\ref{FTfcs}) and (\ref{sympm}).

\medskip
{\it Breakdown} -- 
Interactions can lead to magnetic field asymmetry in nonlinear transport, as
was shown theoretically \cite{sanchez,spivak,polianski06} as well as
experimentally \cite{zumbuhl, wei,   marlow, leturcq, angers}: 
Every particle is moving in a local
potential $U(\vec r)$ generated by all the other particles. 
 The internal
potential  has to be determined  
self-consistently and depends on all potentials $V_\gamma$ applied in the
external contacts, $U(\vec r)=U(\vec r,\{V_\gamma\})$.  
The scattering matrix depends on the energy of the particle and  
is a functional of the internal potential, 
$\mathcal{S}=\mathcal{S}(E,U(\vec r))$. 
Indeed the functional dependence of the scattering matrix 
is required by gauge invariance:
The generating function has to be invariant under a
shift of all potentials by the same amount, $V_\gamma \rightarrow
V_\gamma+U_0$. This condition can be expressed as $\frac{dF}{d U_0}=0$. 
For long times and neglecting interactions, 
the generating function  in the scattering approach is \cite{levitov93}  
\begin{equation}
	F(i{\bf \Lambda}) = \frac{t}{h} \int
	dE\mbox{tr}\left[\ln (\mathbf{1}-\tilde f K)\right]\label{S}.
\end{equation}
Here, $K= (\mathbf{1}-\tilde\lambda^{\dagger}\mathcal{S}^{\dagger} 
	\tilde\lambda  \mathcal{S})$ is composed of the scattering
	matrix 	$\mathcal{S}$, the unit matrix $\mathbf{1}$ and the matrix
	$\tilde\lambda$ introducing the counting fields, 
$\tilde\lambda=\mbox{diag}(e^{-i\lambda_1},e^{-i\lambda_2},\ldots,
e^{-i\lambda_M})$. The diagonal matrix $\tilde f$
	contains the Fermi-functions of the 
	different terminals with $\tilde f=\mbox{diag}(f_1,f_2,\ldots, 
f_M)$. 
With this we can show that  gauge invariance 
requires 
\begin{equation}
	\sum_\gamma \frac{\partial K}{\partial V_\gamma}+e\frac{\partial
	K}{\partial E}=0. 
\end{equation}
Here we used that the derivative of the Fermi-functions with respect to $U_0$ 
can be expressed as an energy-derivative and that 
$\partial K/\partial U_0=\sum_\gamma\partial K/\partial V_\gamma$.  
Therefore, the scattering matrix entering the
expression for $K$ depends not only on the energy $E$ of the carriers
but also via the internal potential landscape on the external voltages 
$\mathcal{S}=\mathcal{S}(E,\{V_\gamma\})$. 
As mentioned above, the local internal potential  has to be determined
 self-consistently  and it is not necessarily an even function of magnetic
 field \cite{sanchez}. As a consequence, for nonlinear transport, the  
 scattering matrix is  not reversible,
 $\mathcal{S}_{\alpha\beta}(B,\{V_\gamma\})\not= 
\mathcal{S}_{\beta\alpha}(-B,\{V_\gamma\})$. This implies 
immediately the  breakdown of the fluctuation theorem (\ref{FTfcs}) and
(\ref{sympm}) for  nonlinear transport, since any derivation is based upon
reciprocity. 

The lack of reversibility can be shown explicitly for the Mach-Zehnder
interferometer. Coulomb-interactions  between the two edge states moving
through the interferometer lead 
to  internal potentials $U_{o}$, $U_{i}$ in the outer and inner edge. In this
respect the inner edge acts as a gate on the outer edge. 
For the interference, this gives rise to 
an additional phase difference $\varphi(B)=e\Delta U_{o}\tau/h$ between the
two interferometer arms. Here, $\tau$ is the 
time an electron needs to traverse the interferometer, and $\Delta U_{o}$
is the difference of the internal potential in the outer edge between upper
and lower arm. 

It is easy to see that the internal screening potential $U_o$ is not an even
function of magnetic field: For positive magnetic field as shown in
Fig.~\ref{mzi}, only processes from left to right contribute, and the internal
potential will depend on the reflection $R_A=1-T_A$ of the left beam splitter
and on the 
voltages $V_1$ and $V_2$. For inversed magnetic field, processes from
right to left are important, which depend on $R_B=1-T_B$ and voltages
$V_3$ and $V_4$. 
To be explicit, we determine the local internal potential self-consistently
within a Hartree approximation \cite{christen,polianski06,sanchez}. 
The average charges  $q_o$ and $q_i$ in the  
edges of the upper interferometer arm are on the one hand expressed as the
difference between injected and screened charge, and are on the other hand
determined by Coulomb interaction. For positive magnetic field, this determines
the internal potentials $U_o$, $U_i$ in the upper arm through
\begin{eqnarray}
	q_i&=&e^2D(V_1-U_{i})=C(U_{i}-U_{o})\label{Qi}\\
	q_o&=&e^2D(R_AV_1+T_AV_2-U_{o})=C(U_{o}-U_{i})\label{Qo}.
\end{eqnarray}
Here, $C$ is the geometric capacitance between the two edges, and $D$ is the
density of states of an edge state. 
Similar equations hold for the lower interferometer arm and for reversed
magnetic field. 
To first order in external voltage, the potential difference $\Delta
  U_{o}=\sum_\alpha u_{\alpha}V_\alpha$ is determined by
the characteristic potentials $u_{\alpha}=[\partial \Delta
  U_{o}/\partial   V_\alpha]_{eq.}$. We find 
$u_{3}(B)=u_{1}(-B)=0$ and  $u_{1}(B)=R_A-e^2DT_A/(2C+e^2D)$,  
$u_{3}(-B)=R_B-e^2DT_B/(2C+e^2D)$, as well as $u_2(B)=-u_1(B)$ and
$u_4(-B)=-u_3(-B)$. 

Using the characteristic 
potentials, the self-consistent transmission probability
$T_{31}=T_{31}(+B,V_1-V_2)$ for a
particle in the interfering edge to transmit from 
terminal $1$ to $3$ for positive magnetic field is
$T_{31}=R_AR_B+T_AT_B-2\sqrt{R_AR_BT_AT_B}\cos(\Phi-\varphi)$ with
$\varphi(+B)=e u_1(+B)\tau(V_1-V_2)/h$.   
For $T_{13}=T_{13}(-B,V_3-V_4)$ at negative magnetic field, the additional 
phase is $\varphi(-B)=e u_3(-B)\tau(V_3-V_4)/h$.  
The lack of reversibility out of equilibrium is evident:  
\begin{equation}
	T_{31}(+B,V_1-V_2)\not=T_{13}(-B,V_3-V_4). 
\end{equation}
This means that the the fluctuation
relation (\ref{sympm}) is strictly speaking valid only at equilibrium but has
corrections for finite voltages. 
In general, taking into account interactions beyond the Hartree-level will
  not reestablish reversibility.

\medskip
{\it Fluctuation relations for correlation functions} -- 
\begin{figure}
\centerline{\psfig{figure=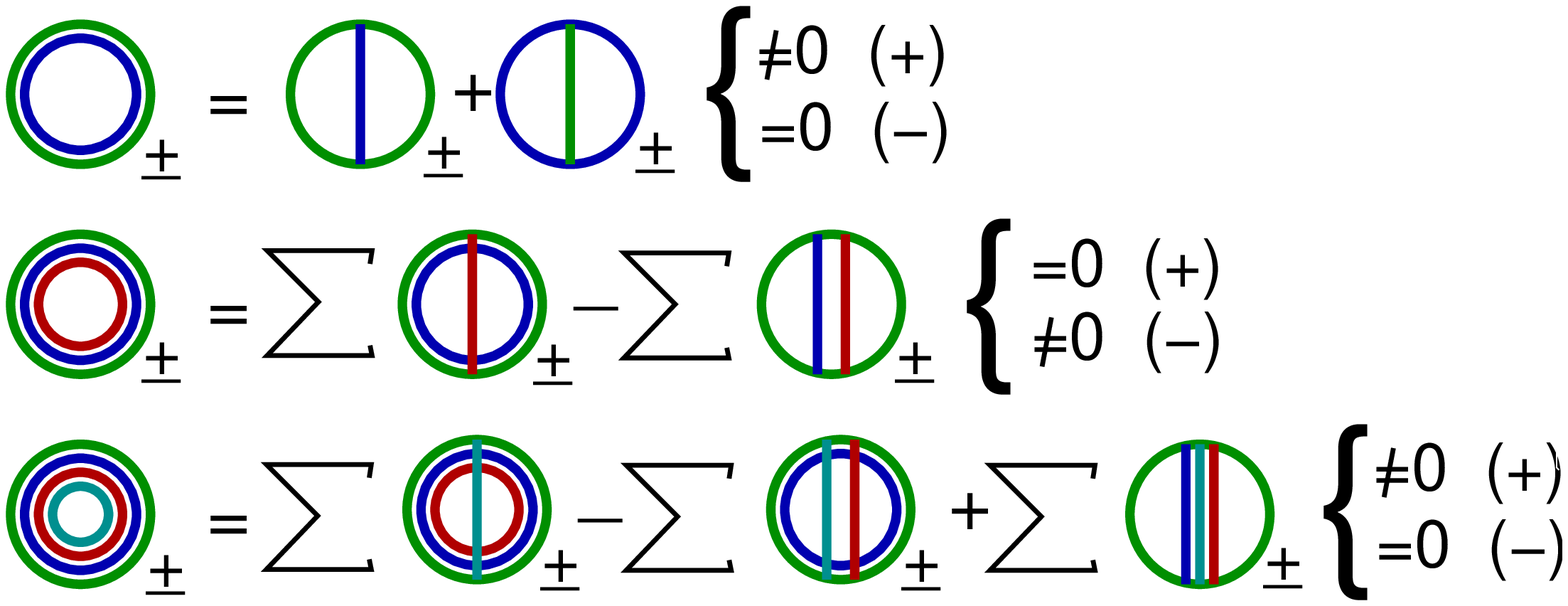,width=8.0cm}}
\caption{(color online) A graphical representation of the relations between
  response 
  coefficients, compare to Eq.~(\ref{relationk}). The number of 
  circles stands for the order of the cumulant, and the number of vertical
  lines  mean the order of the derivative with respect to voltage as well as
  the   power of the factor $(k_BT)$ to which  the
  response    coefficient has to be multiplied. The different heavy lines
  (different colors)   
  represent derivatives with respect to quantities of different terminals. The
  summations go over all possible permutations, where the
  number of permutations is given by ${k\choose l}$ with $k$ the number of
  circles and $l$ the number of vertical lines. Higher order correlations
  follow the same rules and can easily be constructed. }
\label{FTgraphic}
\end{figure}
Including interactions, the fluctuation relation (\ref{sympm}) for the
counting statistics is not valid anymore, as shown above explicitly
  within a   Hartree-model. Nevertheless, we can  
derive fluctuation relations for current correlation functions.  We
  emphasize that the following section is general, no specific model for
  interactions is needed.
It is useful to expand the first few cumulants for $eV \ll k_BT$, 
\begin{eqnarray}
	I_\alpha&=&\sum_\beta G_{\alpha,\beta}^{(1)}V_\beta+\sum_{\beta\gamma}
	G_{\alpha,\beta\gamma}^{(2)}\frac{V_\beta V_\gamma}{2}+\mathcal{O}(V^3)\\
	S_{\alpha\beta}&=&S_{\alpha\beta}^{(0)}+\sum_\gamma
	S_{\alpha\beta,\gamma}^{(1)}V_\gamma+\mathcal{O}(V^2)\\
	C_{\alpha\beta\gamma}&=&C_{\alpha\beta\gamma}^{(0)}+\mathcal{O}(V)
\end{eqnarray}
Up to second order in voltage, the mean current $I_\alpha$ in terminal
$\alpha$ is 
determined by the linear and nonlinear conductance coefficients, $
G_{\alpha,\beta}^{(1)}$ and $G_{\alpha,\beta\gamma}^{(2)}$. The zero-frequency
current
correlations $S_{\alpha\beta}=\av{\Delta I_\alpha\Delta I_\beta}$ contain
equilibrium Nyquist noise $S_{\alpha\beta}^{(0)}$ and the noise susceptibility
 \cite{gabelli} $S_{\alpha\beta,\gamma}^{(1)}$ which includes the emergent shot
noise. Of the 
third cumulant $C_{\alpha\beta\gamma}=\av{\Delta I_\alpha\Delta I_\beta\Delta
  I_\gamma}$, only the equilibrium value $C_{\alpha\beta\gamma}^{(0)}$ is used
in the following. 
All response coefficients are obtained from the
generating function, e.g.~$G^{(2)}_{\alpha,\beta\gamma}=-ie[\partial^3
  F/\partial\lambda_\alpha\partial V_\beta\partial V_\gamma]_0/t$, 
where the index $0$ means
setting ${\bf \Lambda}$ and ${\bf A}$ to zero.

(Anti-)symmetrizing the above definitions, both the
fluctuation-dissipation theorem (for $+$),
and the Onsager-Casimir relations (for $-$),
can be  formulated concisely as  
\begin{equation}\label{cum2}
	S_{\alpha\beta\pm}^{(0)}=k_BT(G_{\alpha,\beta\pm}^{(1)}+G_{\beta,\alpha\pm}^{(1)})=\pm 	S_{\alpha\beta\pm}^{(0)}. 
\end{equation}
The next order fluctuation relation connects the third cumulant at equilibrium
which is odd in magnetic field with combinations of the noise susceptibility
and nonlinear conductance coefficients, 
\begin{eqnarray}
	C_{\alpha\beta\gamma,\pm}^{(0)}&=&k_BT\left(S_{\alpha\beta,\gamma\pm}^{(1)}+S_{\alpha\gamma,\beta\pm}^{(1)}+S_{\beta\gamma,\alpha\pm}^{(1)}\right.\label{C3uni} \\   
	&&\hspace{-10mm}\left.-k_BT(G_{\alpha,\beta\gamma\pm}^{(2)}+G_{\beta,\alpha\gamma\pm}^{(2)}+G_{\gamma,\alpha\beta\pm}^{(2)})\right)=\mp
	C_{\alpha\beta\gamma,\pm}^{(0)} \nonumber
\end{eqnarray}
These universal fluctuation relations can be extended to any order: A current
correlation function at equilibrium is expressed by combinations of successive
response coefficients of lower order current cumulants. They are graphically
represented in figure \ref{FTgraphic}. The first two lines of the figure
correspond  to Eqs.~(\ref{cum2}) and (\ref{C3uni}), higher order relations can
easily be constructed.

The derivation of the fluctuation relations is based on the following
properties of the generating function: 
\begin{eqnarray}
 	F_\pm({\bf -A,A})&=&F_\pm({\bf 0,A})=0\label{F-A}\\
	F_\pm(i{\bf\Lambda,0})&=&\pm F_\pm(-i{\bf\Lambda,0})\label{FT0} 
\end{eqnarray}
The first equation  defines a special symmetry point at  $i{\bf \Lambda=-A}$
for which the generating function vanishes, just as for ${\bf \Lambda=0}$
which originates from probability conservation. To demonstrate it,  
for the case of a non-interacting system, we expand Eq.~(\ref{S}) in
terms of multi-particle scattering events \cite{levitov93} and use the detailed
balance for Fermi functions
$f_\alpha(1-f_\beta)=\exp(A_\alpha-A_\beta)f_\beta(1-f_\alpha)$.  Neither magnetic
field symmetry nor microreversibility are needed.
For a system with electron-electron interactions we start from the 
definition of the generating function  $F(i{\bf \Lambda} )=\ln
\av{e^{-i{\bf   \Lambda \hat Q_t}}e^{i{\bf \Lambda \hat Q_0}}}_0$. 
Here, ${\bf \hat Q_0}$ and ${\bf \hat Q_t}$ denote the charge operators at time
$0$ and time $t$, and the expectation value is taken with respect to the
initial state, described by a grand-canonical density
matrix.  At time $0$ the conductor is
decoupled from the reservoirs, and the initial Hamiltonian $\hat H_0$ commutes
with the charge ${\bf \hat Q_0}$. To derive Eq.~(\ref{F-A}), we use
that the total energy in the system "conductor+reservoirs" is conserved at all
times. For a detailed discussion see appendix B.
We emphasize that the identity Eq.~(\ref{F-A}) is valid without
 microreversibility.  
In terms of distribution functions, Eq.~(\ref{F-A}) defines a global detailed
balance relation,   
\begin{equation}
	\sum_{\bf Q}P({\bf Q})=\sum_{\bf Q}P({\bf Q})e^{-{\bf A Q}}=1,
\end{equation}
 valid
even if Eq.~(\ref{FTfcs}) is not true.
The second equation, Eq.~(\ref{FT0}) represents the fluctuation 
relation (\ref{sympm}) at ${\bf A=0}$ and  is a consequence of microscopic 
reversibility at equilibrium. It follows that even (odd) cumulants at
equilibrium are even (odd) in magnetic field as expressed by: 
\begin{equation}\label{cumeq}
	\av{(\Delta I_\alpha)^k}_\pm^{eq.}=\pm(-1)^k\av{(\Delta
	I_\alpha)^k}_\pm^{eq.}. 
\end{equation}
Both functions $F_\pm(i{\bf\Lambda,A})$ and $F_\pm(-i{\bf\Lambda-A,A})$ can be
expanded as Taylor series around ${\bf A=0}$ and ${\bf \Lambda=0}$. This
defines  general relations for specific Taylor coefficients,
	$\left.\frac{\partial^k F_\pm(-i{\bf\Lambda-A,A})}{\partial
	A_\alpha^k}\right|_0=
	\sum_{l=0}^k {k\choose l}i^l\left.\frac{\partial^k
	F_\pm(i{\bf\Lambda,A})}{\partial
	  A_\alpha^{k-l}\partial\lambda_\alpha^l} \right|_0$,
which vanish identically due to Eq.~(\ref{F-A}). 
The last term in the sum represents the $k$'th derivative of the generating
function with respect to the counting fields, which is the
$k$'th cumulant at equilibrium.
Solving the above equation for this last term leads to
\begin{equation}\label{relationk}
	\av{(\Delta I_\alpha)^k}_\pm^{eq.}=-\sum_{l=1}^{k-1}{k \choose
	l}\left(-k_BT\right)^{k-l} \frac{\partial^{k-l}\av{(\Delta
	I_\alpha)^l}}{\partial 	V_\alpha^{k-l}}\Big|^{eq.}_{\pm}
\end{equation}
This equation relates a correlation function at equilibrium to a linear
combination of response coefficients of lower order correlations. Together with
Eq.~(\ref{cumeq}) -which determines the magnetic field symmetry- it defines
new fluctuation relations for nonlinear transport. In Fig.~\ref{FTgraphic},
they are schematically represented and extended to the general case of a
multi-terminal conductor.

For the MZI, the fluctuation relation Eq.~(\ref{C3uni})
can be explicitly verified within Hartree. We are concerned with temperatures
and voltages low compared to the first plasma mode of an interferometer arm. 
Due to the separation of left- and right-movers, several response coefficients
vanish, in particular the
nonlinear conductance   $G^{(2)}_{3,31\pm}$ as well as
the noise susceptibility $S_{33,1\pm}^{(1)}$. 
Also the third cumulant at equilibrium, $C_{331\pm}^{(0)}$ is zero, because
the  scattering matrix is energy-independent for equal length of the
interferometer arms.  
But the coefficients $G^{(2)}_{1,33}$ and $S^{(1)}_{31,3}$
are finite for $-B$ due to the internal potential, and vanish for
$B$. Similar arguments hold for response coefficients with $1\leftrightarrow
3$.  Using $dg/d U\equiv (4e^3\tau/h^2)\sqrt{R_BT_BR_AT_A}\sin\Phi$, 
Eq.~(\ref{C3uni}) simplifies for the MZI to 
\begin{eqnarray}
	2 S^{(1)}_{31,3\pm}&=&k_BT
	G^{(2)}_{1,33\pm}=\pm k_BT u_3(-B)dg/dU\label{SImzi1}\\ 
	2S^{(1)}_{31,1\pm}&=&k_BTG^{(2)}_{3,11\pm}=
	k_BTu_1(B)dg/dU.\label{SImzi2}
\end{eqnarray}
The fluctuation relation (\ref{sympm}) which does not account for magnetic
field asymmetry in screening effects, would 
require that the anti-symmetrized part ($-$) of the above equations is
identically zero \cite{saito}.  
Measuring a nonlinear conductance coefficient $G^{(2)}_{1,33}$ or noise
susceptibility $S^{(1)}_{31,3}$ which is asymmetric 
in magnetic field proofs Eq.~(\ref{sympm}) wrong. The fluctuation
relations Eqs.~(\ref{SImzi1}) and (\ref{SImzi2}) are linear in temperature, 
periodic with the magnetic flux $\Phi$ and depend on the reflection of the
beam splitters; they can be experimentally
verified.  

\medskip
{\it Conclusion} -- 
We have shown that electron-electron interactions lead to a breakdown of the
usual fluctuation relations for the full counting statistics in the presence of a
magnetic field. The reason is, that 
interactions can induce effective deviations from reversibility of 
scattering processes out of equilibrium. Instead fluctuation
relations can be derived which relate  
correlation functions at equilibrium to response coefficients of correlations
of lower order. These fluctuation relations are valid even in presence of
magnetic field asymmetry.

\medskip
{\it Acknowledgement} --
We thank D. Sanchez and M. Polianski for instructive discussions. 
This work is supported by MaNEP 
and the Swiss NSF. 


\medskip
{\it Appendix A: Derivation of the fluctuation relation} -- 
The fluctuation relation, Eq.~(\ref{FTfcs}), which holds in the presence of
microreversibility, was derived in various systems
\cite{tobiska,andrieux,espositoPRB,saito}, here we demonstrate it for a
non-interacting system described by a scattering matrix. 
Most directly, the symmetry relation Eq.~(\ref{sympm}) in its
non-symmetrized form is derived, $F(i{\bf \Lambda} ,B)=F(-i{\bf \Lambda-A}
,-B)$. The generating function Eq.~(\ref{S})  is
$F(i{\bf \Lambda} )=t\int \frac{dE}{h} H(i{\bf \Lambda})$ with the integrand
\begin{equation}
	H(i{\bf \Lambda} ) =\ln \det
	\left[1+\tilde f\left(\tilde\lambda^{\dagger}\mathcal{S}^{\dagger}
	\tilde\lambda  \mathcal{S}-1\right)\right]\label{H}.
\end{equation}
The determinant can be expanded  in terms of multi-particle scattering
events \cite{levitov93,long}, 
\begin{eqnarray}
	&&H(i{\bf \Lambda} ,B)=\ln \sum_{\{a\},\{b\}}\left| 
	\det\left(\mathcal{S}_{\{b\}}^{\{a\}}(B)\right)\right|^2
	\label{exp}\\ 
	&&	\exp{\Bigl(i\sum_{\alpha\in\{a\}}\lambda_\alpha-i
	\sum_{\alpha\in\{b\}}\lambda_\alpha\Bigr)}
	\prod_{\gamma\in\{a\}}f_\gamma\prod_{\gamma\not\in\{a\}}(1-f_\gamma).
	\nonumber 
\end{eqnarray}
For conductors with a single transport mode,  $\{a\}$ denotes a set of
reservoirs from which particles are 
injected,  and $\{b\}$ is the set of reservoirs into which particles are
emitted. The first sum in Eq.~(\ref{exp}) runs over all possible sets
$\{a\}$ and $\{b\}$, and represents 
all possible, distinct ways of scattering a number $m$ of particles, with $m$
ranging from $0$ to $M$. The probability that  $m$ particles are scattered
from $\{a\}$ to $\{b\}$ is given by
$|\det \mathcal{S}_{\{b\}}^{\{a\}}|^2$, where the matrix
$\mathcal{S}_{\{b\}}^{\{a\}}$ is formed by taking the intersecting matrix
elements of the columns corresponding to
the elements in $\{a\}$ and the rows corresponding to the elements in $\{b\}$
from the scattering matrix $\mathcal{S}$.  
The products over the occupation
functions  of the different terminals in Eq.~(\ref{exp}) determine the
probability that exactly $m$ particles from set $\{a\}$ are injected. 
The exponent contains all counting fields of set $\{a\}$ and $\{b\}$.

Replacing  in Eq.~(\ref{exp}) the magnetic field $B$ by $-B$ and all counting
fields $i\lambda_\alpha$ by $-i\lambda_\alpha-A_\alpha$  leads to additional
exponential factors. 
It is useful to  define the set $\{c\}$ as the contacts from which particles
are injected but
not into which they are emitted, the set $\{d\}$ as those into which particles
are transmitted 
but not from which they are injected, the set $\{e\}$ as the contacts from
which they are
injected and into which they are transmitted, and the set $\{f\}$ as the set
not at all touched.  
Then, with $\{a\}=\{c,e\}$, $\{b\}=\{d,e\}$ and $\not\in\{a\}=\{d,f\}$,
$\not\in\{b\}=\{c,f\}$, the summations and products over elements of the
different sets can be split up, for example
$\prod_{\gamma\in\{a\}}=\prod_{\gamma\in\{c\}}\prod_{\gamma\in\{e\}}$ and
$\sum_{\alpha\in\{b\}}=\sum_{\alpha\in\{d\}}+\sum_{\alpha\in\{c\}}$.
With the help of the detailed balance relation for
Fermi functions, $f_\alpha(1-f_\beta)=e^{A_\alpha-A_\beta}
f_\beta(1-f_\alpha)$, we find for products concerning the  sets $\{c\}$ and
$\{d\}$ 
\begin{eqnarray}
	&&\exp{\Bigl(\sum_{\alpha\in\{d\}}A_\alpha-\sum_{\alpha\in\{c\}}
	A_\alpha\Bigr)}	\prod_{\gamma\in\{c\}}
	f_\gamma\prod_{\gamma\in\{d\}}(1-f_\gamma)=\nonumber\\  
	&&=\prod_{\gamma\in\{c\}}(1-f_\gamma)\prod_{\gamma\in\{d\}}
	f_\gamma.\label{Fermif} 
\end{eqnarray}
The second and crucial point of the derivation is the use of
reciprocity 
$\mathcal{S}_{\alpha\beta}(B)=\mathcal{S}_{\beta\alpha}(-B)$ which implies
$\mathcal{S}_{\{b\}}^{\{a\}}(-B)=\mathcal{S}_{\{a\}}^{\{b\}}(B)$. With this,
and recombining the sets of Fermi functions one obtains finally
\begin{eqnarray}
	&&H({\bf-A}-i{\bf \Lambda},-B)=\ln \sum_{\{a\},\{b\}}\left| 
	\det\left(\mathcal{S}_{\{a\}}^{\{b\}}(B)\right)\right|^2\label{Hmin2}\\
	&&\exp{\Bigl(i\sum_{\alpha\in\{b\}}\lambda_\alpha- 
	i\sum_{\alpha\in\{a\}}\lambda_\alpha\Bigr)}
	\prod_{\gamma\in\{b\}}f_\gamma\prod_{\gamma\not\in\{b\}}(1-f_\gamma),
	\nonumber
\end{eqnarray}
which is indeed equal to $H(i{\bf \Lambda},B)$ in Eq.~(\ref{exp}), since the
sum runs over all possible sets $\{a\}$ and $\{b\}$.

As discussed in the core of the paper, screening effects lead to an
internal potential which can be asymmetric in magnetic field. Then, away from
equilibrium, the reciprocity relation is not valid,  
$\mathcal{S}_{\alpha\beta}(E,U(B))\not=\mathcal{S}_{\beta\alpha}(E,U(-B))$,
and with this the symmetry relation (\ref{sympm}) breaks down.


\medskip
{\it Appendix B: Symmetry point of the generating function} --
Eq.~(\ref{F-A}) defines a symmetry point for the generating function which is
valid for any value of the  magnetic field, $F({\bf -A},B)=F({\bf 0},B)=0$.
It can be derived without the use of microreversibility. For a system,
described by a scattering matrix, we start with Eq.~(\ref{exp}) at ${i\bf
  \Lambda =-A}$ 
\begin{eqnarray}
	&&H({\bf -A})=\ln \sum_{\{a\},\{b\}}\left| 
	\det\left(\mathcal{S}_{\{b\}}^{\{a\}}\right)\right|^2\\
	&&\exp{\Bigl(-\sum_{\alpha\in\{a\}}A_\alpha+\sum_{\alpha\in\{b\}}
	A_\alpha\Bigr)}
	\prod_{\gamma\in\{a\}}f_\gamma\prod_{\gamma\not\in\{a\}}(1-f_\gamma).
	\nonumber  
\end{eqnarray}
Reformulating the sets of contacts $\{a\}$ and $\{b\}$ in terms of
$\{c\}-\{f\}$ as introduced 
above and inserting the property (\ref{Fermif}), the exponential factors will
be absorbed into the Fermi functions and we find
\begin{eqnarray}
	H({\bf-A})&=&\ln \sum_{\{a\},\{b\}}\left| 
	\det\left(\mathcal{S}_{\{a\}}^{\{b\}}\right)\right|^2
	\prod_{\gamma\in\{b\}}f_\gamma\prod_{\gamma\not\in\{b\}}(1-f_\gamma)
	\nonumber\\ 	
	&=&\ln	\sum_{\{b\}}\left(\prod_{\gamma\in\{b\}}
	f_\gamma\prod_{\gamma\not\in\{b\}}(1-f_\gamma)\right)=0 
\end{eqnarray}
In the first line, the sum over $\{a\}$ can be performed and equals one
because of probability 
conservation. In the scattering picture it is thus easy to see that the
identity $F_\pm({\bf 0})=F_\pm({\bf -A})=0$ is a consequence of both
  probability conservation and the detailed balance for Fermi functions. 

For  systems with arbitrary interactions, the generating
function for charge transfer 
is defined as an expectation value with respect to the initial state \cite{lll}
\begin{equation}
	F(i{\bf \Lambda} )=\ln \av{e^{-i{\bf \Lambda \hat Q_t}}e^{i{\bf \Lambda
	\hat Q_0}}}_0. 
\end{equation}
Here, the vector quantities ${\bf \hat Q_0}$ and ${\bf \hat Q_t}$ denote the
charge operators in the different terminals at time 
$0$ and time $t$. The initial state is described by the grand-canonical
density matrix $\hat \rho_0=e^{-\beta \hat H_0+{\bf A \hat
    Q_0}}/Z_0$, with the partition sum $Z_0=tr[e^{-\beta\hat H_0+{\bf A\hat
      Q_0}}]$.   The Hamiltonian $\hat H_0$ is composed of the Hamiltonians of
all reservoirs and of the scatterer, which at time $0$ are
supposed to be decoupled.  Importantly, the time evolution operator $\hat
U(t)$ contains in addition the coupling to the reservoir and interaction terms
and does not commute with $\hat H_0$.
Inserting the initial density matrix into the definition above  and using the
cyclic property of the trace as well as the fact that $\hat H_0$ and ${\bf
  \hat Q_0}$ commute, we  obtain 
\begin{eqnarray}
	F({\bf -A})&=&\ln \av{e^{{\bf A \hat Q_t}}e^{{\bf -A\hat
 	Q_0}}}_0=\label{trace}\\ 
	&=&\ln\frac{1}{Z_0}tr\left[e^{-\beta\hat H_0+{\bf A \hat
	Q_0}}\hat U^\dagger(t) 
	e^{{\bf A \hat Q_0}}\hat U(t)e^{-{\bf A \hat Q_0}}\right]\nonumber\\
	 &=&\ln\frac{1}{Z_0}tr\left[e^{-\beta\hat H_0}\hat U^\dagger(t)
	e^{{\bf A \hat Q_0}}\hat U(t)\right]=\nonumber\\	
	&=&\ln\frac{1}{Z_0}tr\left[e^{-\beta\hat H_0+{\bf A\hat Q_0}}
	e^{\beta\hat H_0}\hat U(t) e^{-\beta\hat H_0}\hat
	U^\dagger(t)\right].\nonumber 
\end{eqnarray}
The trace is evaluated by $tr[\ldots]=\sum_n\av{n|\ldots|n}$, where we choose
the eigenbasis of the initial Hamiltonian,
$H_0|n\rangle=\epsilon_n(0)|n\rangle$.  The state $|n\rangle$ is characterized
by a configuration of numbers of particles in all reservoirs and the
scatterer. We introduce the total energy  $\epsilon_n(0)$ of this 
particular configuration. The important point we can make is that scattering
processes through the conductor leave this energy invariant.
To proceed, we consider a matrix element of the last four operators in the
last line of 
Eqs.~(\ref{trace}), insert $1=\sum_k|k\rangle\langle k|$ in the middle and find
\begin{eqnarray}
	&&\hspace{-1.7cm}\langle m|e^{\beta\hat H_0}\hat U(t) e^{-\beta\hat
	H_0}\hat U^\dagger(t)|n\rangle=\label{matrixel}\\
        &&=\sum_k \langle m|e^{\beta\hat
	H_0}|k(t)\rangle e^{-\beta\epsilon_k(0)}\langle
	k(t)|m\rangle\nonumber\\  
        &&=\sum_k \langle m|k(t)\rangle \langle k(t)|n\rangle=\delta_{nm}.\nonumber 
\end{eqnarray}
For the last step we used that
$H_0|k(t)\rangle=\epsilon_k(t)|k(t)\rangle$. This is justified, since
the time evolution of state $|k\rangle$ simply means a change of the numbers
of particles in each reservoir. Assuming that scattering processes are
instantaneous, it  will at any time form a basis for $\hat H_0$.  Then, using
that the total energy is 
conserved, $\epsilon_k(t)=\epsilon_k(0)$, the matrix element (\ref{matrixel})
is diagonal.  
This result leads to $F({\bf -A})=\ln tr[e^{-\beta\hat
    H_0+{\bf A\hat     Q_0}}]/Z_0=0$ as   in Eq.~(\ref{F-A}).
We emphasize again that the identity Eq.~(\ref{F-A}) is valid for arbitrary
electron-electron interactions and without the use of microreversibility. 
Special care should be taken of the case when (i) the problem is
 time-dependent, (ii) the temperature is not equal
in all reservoirs, and (iii) a bath allows energy exchange, e.g.~via
electron-phonon interactions. Then, we would have to consider energy currents
as well and  introduce additional counting fields that account for the
transferred energy. In this case, a similar relation can be derived.

\end{document}